\documentclass[twocolumn,numberedappendix]{emulateapj}
\bibpunct[; ]{(}{)}{;}{a}{}{;}
\usepackage{apjfonts}
\usepackage{graphicx}
\usepackage{natbib}
\usepackage{amsmath}
\usepackage{upgreek}

\newcommand{\Msolar}{M${_\odot}$}

\shorttitle{Anomalous extinction towards NGC\,1938}
\shortauthors{De Marchi, Panagia, Milone}

\received{2 June 2020}
\accepted{20 July 2020}

\begin{document}

\title{Anomalous extinction towards NGC\,1938\,\altaffilmark{*}}

\author{
Guido De Marchi,\altaffilmark{1}
Nino Panagia,\altaffilmark{2,3}
Antonino P. Milone\,\altaffilmark{4,5}
}

\altaffiltext{1}{European Space Research and Technology Centre,
Keplerlaan 1, 2200 AG Noordwijk, Netherlands; gdemarchi@esa.int}
\altaffiltext{2}{Space Telescope Science Institute, 3700 San Martin
Drive, Baltimore MD 21218, USA; panagia@stsci.edu}
\altaffiltext{3}{Supernova Limited, OYV \#131, Northsound Rd., Virgin Gorda
VG1150, Virgin Islands, UK}
\altaffiltext{4}{Dipartimento di Fisica e Astronomia, 
Univ. di Padova, Vicolo dell'Osservatorio 3, Padova I-35122, Italy; 
antonino.milone@unipd.it}
\altaffiltext{5}{Istituto Nazionale di Astrofisica -- Osservatorio 
Astronomico di Padova, Vicolo dell'Osservatorio 5, Padova I-35122,
Italy}

\altaffiltext{{$\star$}}{Based on observations with the NASA/ESA {\it Hubble 
Space Telescope}, obtained at the Space Telescope Science Institute, which 
is operated by AURA, Inc., under NASA contract NAS5-26555}

\begin{abstract}  

Intrigued by the extended red-giant clump (RC) stretching across the
colour--magnitude diagram of the stars in a $50\times50$\,pc$^2$ region
of the Large Magellanic Cloud (LMC) containing the clusters NGC\,1938
and NGC\,1939, we have studied the stellar populations to learn about
the properties of the interstellar medium (ISM) in this area. The
extended RC is caused by a large and uneven amount of extinction across
the field. Its slope reveals anomalous extinction properties, with
$A_V/E(B-V) \simeq 4.3$, indicating the presence of an additional grey
component in the optical contributing about 30\,\% of the total
extinction  in the field and requiring big grains to be about twice as
abundant as in the diffuse ISM. This appears to be consistent
with the amount of big grains injected into the surrounding ISM by the
about 70 SNII explosions estimated to have occurred during the lifetime
of the $\sim 120$\,Myr old NGC\,1938. Although this cluster appears
today relatively small and would be hard to detect beyond the distance
of M\,31, with an estimated initial mass of $\sim 4\,800$\,\Msolar\
NGC\,1938 appears to have seriously altered the extinction properties in
a wide area. This has important implications for the interpretation of
luminosities and masses of star-forming galaxies, both nearby and in the
early universe. 

\end{abstract}


\section{Introduction}

NCG\,1938 and NGC\,1939 are two clusters in the Large Magellanic Cloud
(LMC), about $36\arcsec$ apart and with a projected separation of 9\,pc
at the distance of the LMC ($51.4 \pm 1.2$\,kpc; Panagia et al. 1991,
and updates in Panagia 1998, 1999). They are projected against the
central portion of the galaxy, about $10^\prime$ south of the LMC bar,
in a region of high stellar density. We owe the first photometric study
of their stellar populations to Mackey \& Gilmore (2004), who observed
the clusters with the  {\em Hubble Space Telescope} (HST). These authors
showed that NGC\,1939 is an old globular cluster (GC) with age and
metallicity comparable to those of the metal-poor Galactic GCs,
confirming the earlier analysis of integrated spectroscopic data of this
cluster by Dutra et al. (1999; see also Piatti et al. 2018). As for
NGC\,1938, from a coarse colour--magnitude diagram (CMD) Mackey \&
Gilmore (2004) estimated an approximate upper limit of 400 Myr to its
age. The pronounced age difference between the two clusters suggests
that they are not physically related and simply appear close due to a
fortuitous projection effect.

A remarkable feature in the CMD of this field (see Figure\,2 in Mackey
\& Gilmore 2004) is the prominent extension of the red-giant clump (RC),
which clearly indicates that in this field there is a considerable
amount of patchy extinction. De Marchi et al. (2014, 2016) and De Marchi
\& Panagia (2014, 2019) have shown how in an extragalactic environment
with a planar geometry such as the LMC, where all stars are practically
at the same distance from us (to within a few percent, see e.g. van der
Marel \& Cioni 2001), the extended RC feature can be used to measure the
direction of the reddening vector in the CMD, the absolute value of the
extinction towards each RC star, the extinction law when multi-colour
photometry is available, as well as information on the distribution and
abundance of grains in the interstellar medium (ISM).  

In this work, we investigate the origin of the patchy extinction causing
differential reddening across the field around NGC\,1938 and NGC\,1939
and show that the extinction properties and the grain size distribution that
they imply appear to be consistent with the number and mass of big
grains that explosions of supernovae of type II (SNII) in the NGC\,1938
cluster have injected into the ISM over the past $\sim 120$\,Myr. 

The structure of the paper is as follows. In Section\,2 we present the
observations and their analysis. In Section\,3 we discuss the different
populations present in the field, while Section\,4 is devoted to the
extinction properties, which we compare with those of other regions in
the LMC. A discussion and the conclusions follow, respectively, in
Sections\,5 and 6. 

\section{Observations and data analysis}

\begin{figure}
\centering
\includegraphics[clip,trim=3.6cm 0cm 3.6cm 0cm, width=0.5\textwidth]{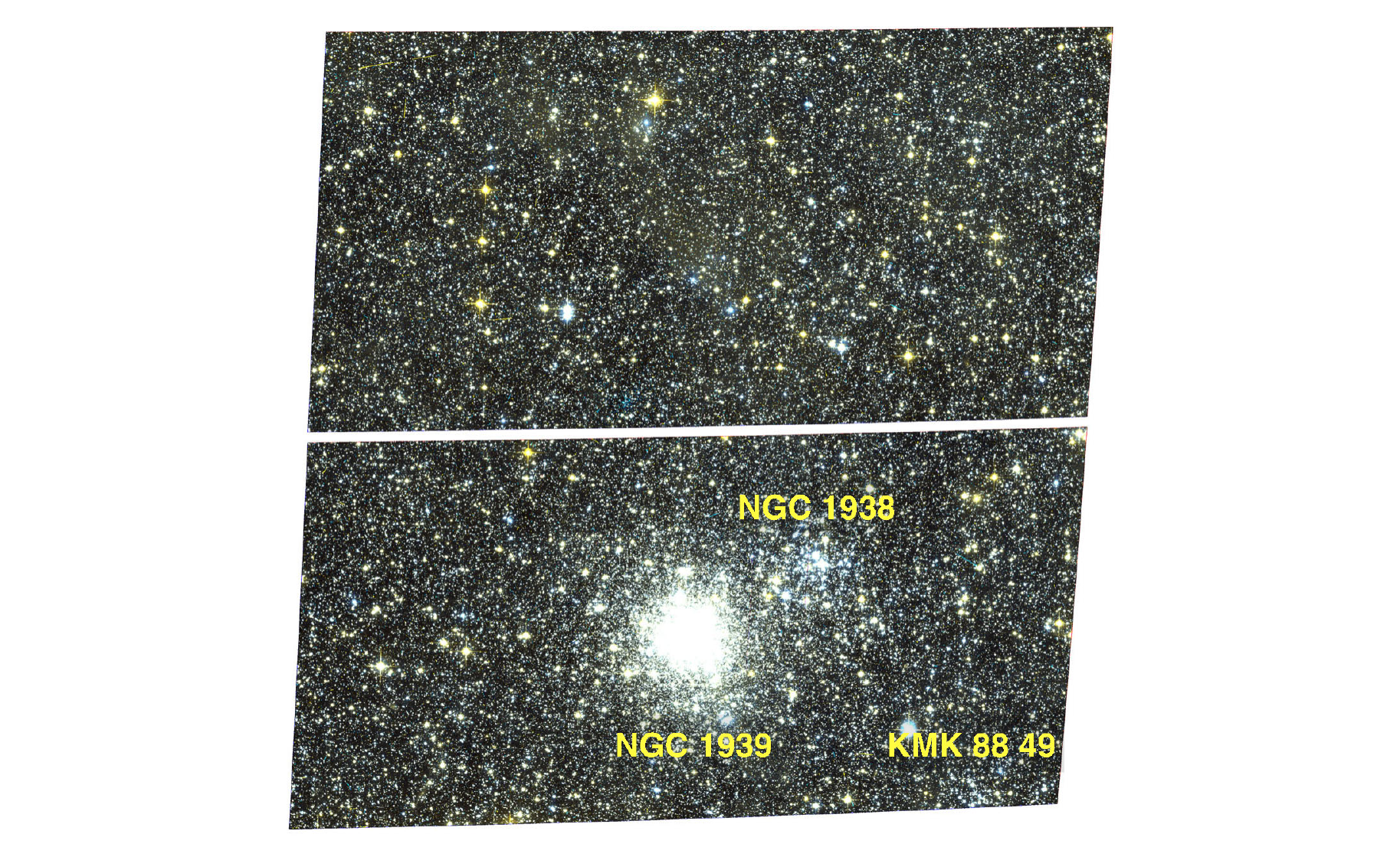}
\caption{Colour composite image of the region obtained by combining the
exposures in the {F555W and F814W} bands. The field spans approximately
$205\arcsec$ or $\sim 50$\,pc on a side. North is inclined $44^\circ$
to the right of the vertical, with East to the left of North.}
\label{fig1}
\end{figure}

The clusters NGC\,1938 and NGC\,1939 were observed on 2003 July 27 with
the Wide Field Channel (WFC) instrument of the Advanced Camera for
Surveys on board the HST (proposal 9891). Two exposures were collected,
one of 330\,s through the F555W filter and one of 200\,s through the
F814W filter. {These filters are rather similar to the standard 
Johnson $V$ and $I$ bands, although they are not identical.}
We retrieved the fully calibrated, flat-fielded science-ready data
corrected for the effects of poor charge transfer efficiency from the
Mikulski Archive for Space Telescopes. A colour-composite image
obtained by combining the two exposures is shown in Figure\,\ref{fig1}.
We note in passing that this field contains also parts of a third and
rather small cluster, namely KMK88\,49 (Kontizas et al. 1988; Piatti
2017), located near the lower right corner in Figure\,\ref{fig1}.

The astrometric and photometric analysis was carried out following the
effective point spread function (ePSF) fitting procedure developed by
Anderson et al. (2008) and stellar positions were corrected for
geometric distortion by using the solution by Anderson \& King (2006).
With the exposure times used for these observations, the  detector's 
response for stars brighter than $m_{555} \simeq 18.1$ and
$m_{814}\simeq 17.4$ becomes progressively non linear and saturates.
These stars' brightness was however completely recovered by summing over
pixels into which bleeding occurred as a result of the over-saturation
(see Gilliland 2004 and Anderson et al. 2008 for details). Magnitudes
were calibrated in the VEGAMAG reference system following Anderson et
al. (2008) and using the most recent zero-point values available through
the ACS Zeropoints Calculator (see Ryon 2019). Photometric uncertainties
are very small, ranging from $0.01$\,mag at $m_{555}=18.2$ to $0.1$\,mag
at $m_{555}\simeq 25$ (details are provided in Table\,\ref{tab1}). The
magnitudes of stars within about 2\,mag of the saturation limit  are
recovered with an accuracy of typically $0.02-0.03$\,mag.

\begin{deluxetable}{lll} \tablecolumns{3}
\tabletypesize{\footnotesize}
\tablecaption{Photometric uncertainties.
\label{tab1}}
\tablewidth{9cm}
\tablehead{\colhead{$m_{555}$} & \colhead{$\sigma_{555}$} &
\colhead{$\sigma_{555-814}$} \\[0.05cm]
\multicolumn{1}{c}{(1)} & \multicolumn{1}{c}{(2)} &
\multicolumn{1}{c}{(3)}}
\startdata
  $18.20$ &  $0.010$ & $0.014$ \\
  $18.70$ &  $0.011$ & $0.015$ \\
  $19.20$ &  $0.011$ & $0.015$ \\
  $19.70$ &  $0.012$ & $0.017$ \\
  $20.20$ &  $0.013$ & $0.018$ \\
  $20.70$ &  $0.013$ & $0.019$ \\
  $21.20$ &  $0.015$ & $0.022$ \\
  $21.70$ &  $0.017$ & $0.026$ \\
  $22.20$ &  $0.020$ & $0.034$ \\
  $22.70$ &  $0.027$ & $0.045$ \\
  $23.20$ &  $0.041$ & $0.068$ \\
  $23.70$ &  $0.054$ & $0.090$ \\
  $24.20$ &  $0.078$ & $0.126$ \\
  $24.70$ &  $0.098$ & $0.142$ \\
  $25.20$ &  $0.130$ & $0.176$ 
\enddata
\tablecomments{Table columns are as follows: (1) $m_{555}$ magnitude;
(2) uncertainty on the $m_{555}$ magnitude; (3) uncertainty on the
$m_{555}-m_{814}$ colour.}
\end{deluxetable}

We included in our analysis only stars that are well fitted by the ePSF
routine (Anderson et al. 2008) and have small root-mean-square scatter
in position measurements. This sample of stars with ``high quality''
photometry was selected as in Milone et al. (2009, see their Figure\.1)
on the basis of the various diagnostics of the astrometric and
photometric quality provided by the computer programmes by 
Anderson et al. (2008).

\begin{figure*}
\centering
\resizebox{\hsize}{!}{\includegraphics{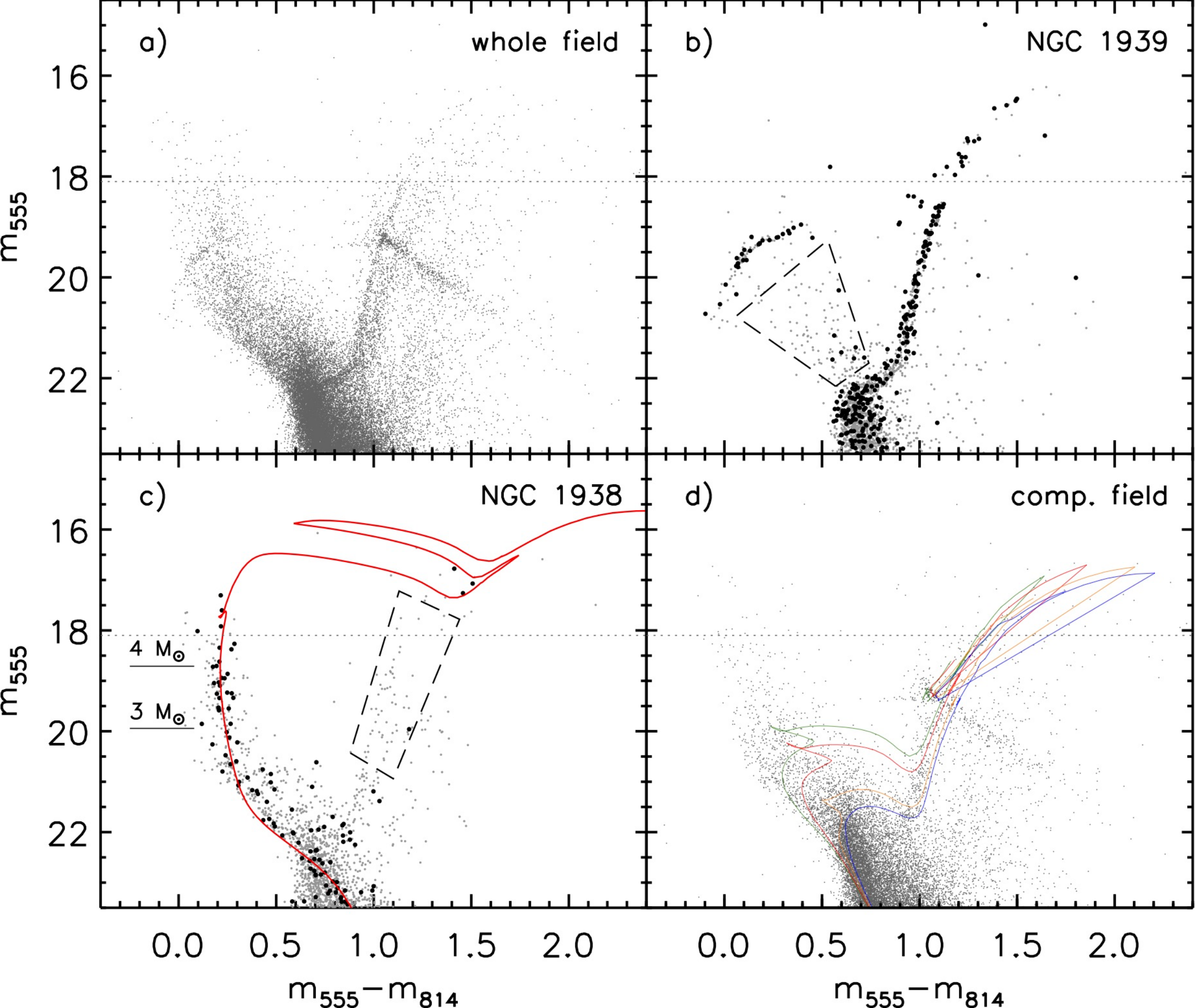}}
\caption{CMDs of high-quality stars in different parts of the field. In Panel 
a) all high-quality stars are included. In Panel b), small grey dots are 
used for stars within a radius of $5.5$\,pc of the centre of NGC\,1939, 
while thick black dots mark objects within $2.5$\,pc of it. In Panel c), 
small grey dots indicate stars within $5.5$\,pc of the centre of NGC\,1938 
and thick black dots those within a $1.6$\,pc radius. The solid line shows a
120\,Myr isochrone from the models of Marigo et al. (2008) for the 
appropriate metallicity and distance, and combined colour excess
(foreground + intrinsic) of $E(V-I)=0.39$. The CMD in  Panel d) is
obtained from all stars contained in the top half of Figure\,\ref{fig1}.
Theoretical isochrones from the models of Marigo et al. (2008) for
metallicity $Z=0.004$ and ages of $1.5$, 2, 3,  and 4\,Gyr are shown,
respectively, in green, red, orange, and blue. In all panels the thin
horizontal lines indicate the saturation level discussed in Section\,2.}  
\label{fig2}
\end{figure*}

\section{Colour--magnitude diagrams of different populations}

We show in Figure\,\ref{fig2}a the CMD obtained from all the stars with
high-quality photometry in this field. The CMD reveals a complex
population, made up of stars in different evolutionary phases: a
prominent and broad  main sequence (MS), a double red giant branch
(RGB), and a remarkably elongated RC, extending by over one magnitude in
$m_{555}$. The broadened MS and extended RC suggest differential reddening
caused by patchy absorption across the field, as Mackey \& Gilmore
(2004) already pointed out.

The majority of the stars in this area are LMC field objects, with a
smaller contribution coming from the NGC\,1938 and NGC\,1939 cluster
themselves. To better characterise the individual components of the
overall stellar population in this area, we display in the other panels
of Figure\,\ref{fig2} the CMDs of different portions of this region. 

In Figure\,\ref{fig2}b, small grey dots correspond to stars with
high-quality photometry within a radius of $22\arcsec$ or $5.5$\,pc
around the nominal centre of NGC\,1939. To limit the weight of field
stars and secure a higher density of cluster members, we show as thicker
black dots the objects within $10\arcsec$ or $2.5$\,pc of the cluster
centre. The resultant CMD reveals a well defined MS turn-off, at
$m_{555} \simeq 22.5$, a narrow and sharp RGB, extending to
$m_{555}\simeq 16$, as well as a prominent blue horizontal branch (HB),
reaching down to $m_{555} \simeq 21$. As Mackey \& Gilmore (2004)
pointed out, the CMD of NGC\,1939 resembles very closely those of
metal-poor Galactic GCs such as for instance M\,15 or M\,92. Most of the
objects located between the MS turn-off and the blue HB, which could in
principle be blue stragglers (e.g. Sandage 1953; Ferraro et. al 2003),
are in fact statistically compatible with being all field stars (there
are 91 objects inside the wedge in Figure\,\ref{fig2}b and about 110
objects in an equally sized region located elsewhere at the NE edge of
the field).

From a detailed study of the HB at RGB, Mackey \& Gilmore (2004) derived
for NGC\,1939 a distance modulus $(m-M)_0=18.48 \pm 0.16$ and a colour
excess $E(V-I)=0.16$, corresponding in turn to $E(B-V)=0.11$ and
$A_V=0.34$ adopting the Galactic extinction law for the diffuse ISM
(e.g. Savage \& Mathis 1979). Although this is slightly higher than the
canonical $E(B-V)=0.07$ value usually adopted for the Galactic
foreground contribution to the extinction towards the LMC (e.g.
Fitzpatrick \& Savage 1984), the colour excess derived by Mackey \&
Gilmore (2004) is still within the range of literature values (see, e.g.
McNamara \& Feltz 1980). It is interesting to note, however, that none
of the observed sequences in the CMD of NGC\,1939 appear to be broadened
by the differential reddening present in the field. This promptly
suggests that the NGC\,1939 cluster is in the foreground of the LMC and
that something else along the line of sight beyond the cluster itself is
causing patchy extinction and differential reddening. {Indeed, that
NGC\,1939 could be an outer disc cluster projected onto the LMC bar had
already been suggested by Piatti et al. (2019). These authors derived a
height of 450\,pc out of the LMC plane based on the combined analysis of
their radial velocity measurements with proper motion  observations from
{\em Gaia}.}

As concerns extinction towards NGC\,1938, the situation is rather
different. In Figure\,\ref{fig2}c we show the CMD of all the stars with
high-quality photometry in this cluster. Small grey dots mark stars
within $5.5$\,pc of the nominal cluster centre, while the thick black
dots are the more centrally located objects within a radius of
$6\farcs5$ or $1.6$\,pc. The main feature in this CMD is a prominent
upper MS reaching $m_{555}\simeq17$, witnessing a relatively young stellar
population. This is confirmed by a sparsely populated RGB in the central
regions of the cluster, which is consistent with field contamination
(there are 56 objects inside the dashed wedge in Figure\,\ref{fig2}c and
about 44 objects in an equally sized sky area located elsewhere at the
same distance from the NGC\,1939 cluster).

The upper MS of NGC\,1938 appears broad, witnessing the effects of
patchy extinction. Furthermore, also the mean extinction towards
NGC\,1938 appears considerably higher than the $E(V-I)=0.16$ value
measured by Mackey \& Gilmore for NGC\,1939. We can satisfactorily
reproduce the NGC\,1938 young MS and what appear to be red supergiants
with a theoretical isochrone from Marigo et al. (2008) for an age of
120\,Myr, metallicity $Z=0.007$, and adopting for the LMC a distance
modulus $(m-M)_0=18.55$ (Panagia 1998). However, to obtain a good fit an
extra $E(V-I)=0.23$ component of colour excess is needed in addition to
the foreground $E(V-I)=0.16$ mentioned above. If the more canonical
$E(V-I)=0.11$ is used for the foreground extinction to the LMC, the
contribution of the extra component grows to $E(V-I)=0.28$ with no
noticeable difference in the quality of the isochrone fit.

As we will show in Section\,4, also the value of $R_V$ in NGC\,1938 is
larger than the canonical $R_V=3.1$ characteristic of the diffuse ISM.
We note that isochrones for younger or older ages would give a
progressively less good fit. The only existing estimate of the age of
NGC\,1938 is that of Mackey \& Gilmore (2004), who set an upper limit of
400\,Myr to it. Our isochrone best fit, however, indicates a definitely
younger age for this cluster, of the order of 120\,Myr. The isochrone
suggests that the heaviest stars still on the MS have a mass of
approximately $4.5$\,\Msolar. The emerging picture is one in which
NGC\,1938 is considerably younger and more extinguished than NGC\,1939
and likely closer to the LMC than the latter, most likely inside
the LMC disc itself. 

To compare this situation with that of the surrounding field, in
Figure\,\ref{fig2}d we show the CMD obtained from all the stars with
high-quality photometry falling in the ACS detector not including the
two clusters (top half of Figure\,\ref{fig1}). The CMD shows the
structures typical of the LMC field population, including a MS, RGB, and
RC, which in this field all appear to be severely affected by
differential reddening. To guide the eye, we show theoretical isochrones
for ages of $1.5$, 2, 3, and 4\,Gyr from the same models of Marigo et
al. (2008), shown respectively in green, red, orange, and blue.
Measurements reported in the literature concur in assigning ages of
$1-3$\,Gyr to the intermediate-age populations in the field of the LMC 
(e.g., Westerlund et al. 1995; Elson et al. 1997; Geha et al. 1998), and
evidence  points to a major star formation event occurred some 2\,Gyr
ago. As for the metallicity, we adopt as representative for this older
stellar population $Z=0.004$, which is the preferred value for field
stars in 30\,Dor (De Marchi et al. 2014) and is typical for the LMC
(e.g., Hill et al. 1995; Geha et al. 1998). 

Adopting for the foreground component of the extinction the
$E(V-I)=0.16$ value derived by Mackey \& Gilmore (2004) for NGC\,1939,
the isochrones provide a good fit to the  RGB and to the location of the
`head' of the RC at $m_{555}-m_{814} \simeq 1.1$ and $m_{555} \simeq
19.3$. Nevertheless, the overdensity of points in the CMD departing from
the head of the RC and extending by more than one magnitude in $m_{555}$
is a clear and unmistakable sign of patchy extinction. In the next
section we will address the use of this feature to derive the direction
of the reddening vector from the data themselves.

We note that the CMD of KMK88\,49, not shown in Figure\,\ref{fig2}, is
marginally different from that of field stars in the same area. From the
analysis of its integrated colours, Piatti (2017) estimated an age of
$\ga 500$\,Myr, while the total mass is estimated by Popescu et al.
(2012) to be around 800\,\Msolar, also from integrated colours.

\section{Red clump stars as extinction probes}

The RC is populated by stars that are experiencing their central
He-burning phase (e.g. Cannon 1970). The remarkably consistent location
of these stars in the CMD makes the RC feature a powerful tool to probe 
distance and reddening. A detailed study of the behaviour of the mean RC
as a function of age, metallicity, and star-formation history is
provided by Girardi \& Salaris (2001) and Salaris \& Girardi (2002).
Populations containing RC stars with different age and metallicity will
produce a slightly elongated shape for the RC, but, as shown by De
Marchi et al. (2014), in the LMC these effects account for at most
$0.2$\,mag in the F555W and F814W bands when both the metallicity and age
vary by a factor of two. 

Also differences in distance between RC members in the same field could
cause the RC to extend in the CMD, as is commonly observed for field RC
stars in the Milky Way. But in order to cause a one magnitude extension,
such as the one observed in Figure\,\ref{fig2}, the RC objects at the
most distant end must be $\sqrt{2.5}$ times farther away than those at
the near end. In the case of the LMC, this {would require the galaxy
to extend about 30\,kpc along the line of sight and is at odds with the
accepted almost planar geometry. Van der Marel \& Cioni (2001) derived
an inclination angle of $\sim 35^\circ$ for the plane of the LMC and a
disc scale height of $\la 0.5$\,kpc. A somewhat larger estimate of the
scale height comes from the recent work of Jacyszyn--Dobrzeniecka et al.
(2016), who probed the three-dimensional structure using classical
Cepheids. Their data show that, when considering the entire extent of
the LMC, the standard deviation on Cepheids distances is about 2\,kpc,
due to the inclination of the galaxy with respect to the plane of the
sky, but it becomes obviously smaller when individual lines of sight are
considered. For instance, within} a radius of 50\,pc (200$\arcsec$) of
the centre of our region, the catalogue of  Jacyszyn--Dobrzeniecka et
al. (2016) contains six classical Cepheids, with a median distance of
$51.4$\,kpc and a standard deviation of $0.75$\,kpc, implying a scale
height of the order of $\sim 1$\,kpc. {Even in this case, without
differential reddening the expected elongation of the RC would still be
less than $0.1$\,mag. So the morphology of the RC observed in
Figure\,\ref{fig2}d is not consistent with any physically reasonable
extension of the LMC along the line of sight and can only be the result
of high and patchy extinction.}

In these circumstances, the RC shape in the CMD can be used to measure 
the direction of the reddening vector in a fully empirical way (Nataf et
al. 2013; De Marchi et al. 2014). An efficient and accurate method to
achieve this is the application of the unsharp-masking technique.
Extensive details on the method are provided by De Marchi et al. (2016).
In summary, unsharp-masking makes an image of the CMD sharper by
overlapping to the image itself a mask consisting of an inverted blurred
version of the image. To this aim, each object in the CMD is mapped to a
two-dimensional array (the CMD image)  with a sampling of $0.01$ mag in
colour and magnitude and the array is then convolved with a narrow
Gaussian beam. The convolution assigns to each point in the CMD the
proper resolution, including uncertainties on the photometry, and here
we used $\sigma=0.08$\,mag, or about three times the typical photometric
uncertainty. To create the mask, we convolved the array also with a
wider Gaussian beam, with $\sigma=0.3$\,mag, and subtracted the mask
from the CMD image. These operations are analytically equivalent to
convolving the CMD with a kernel represented by the difference between
two Gaussian beams with different $\sigma$ (see De Marchi et al. 2016
for further details).

\begin{figure}
\centering
\resizebox{\hsize}{!}{\includegraphics{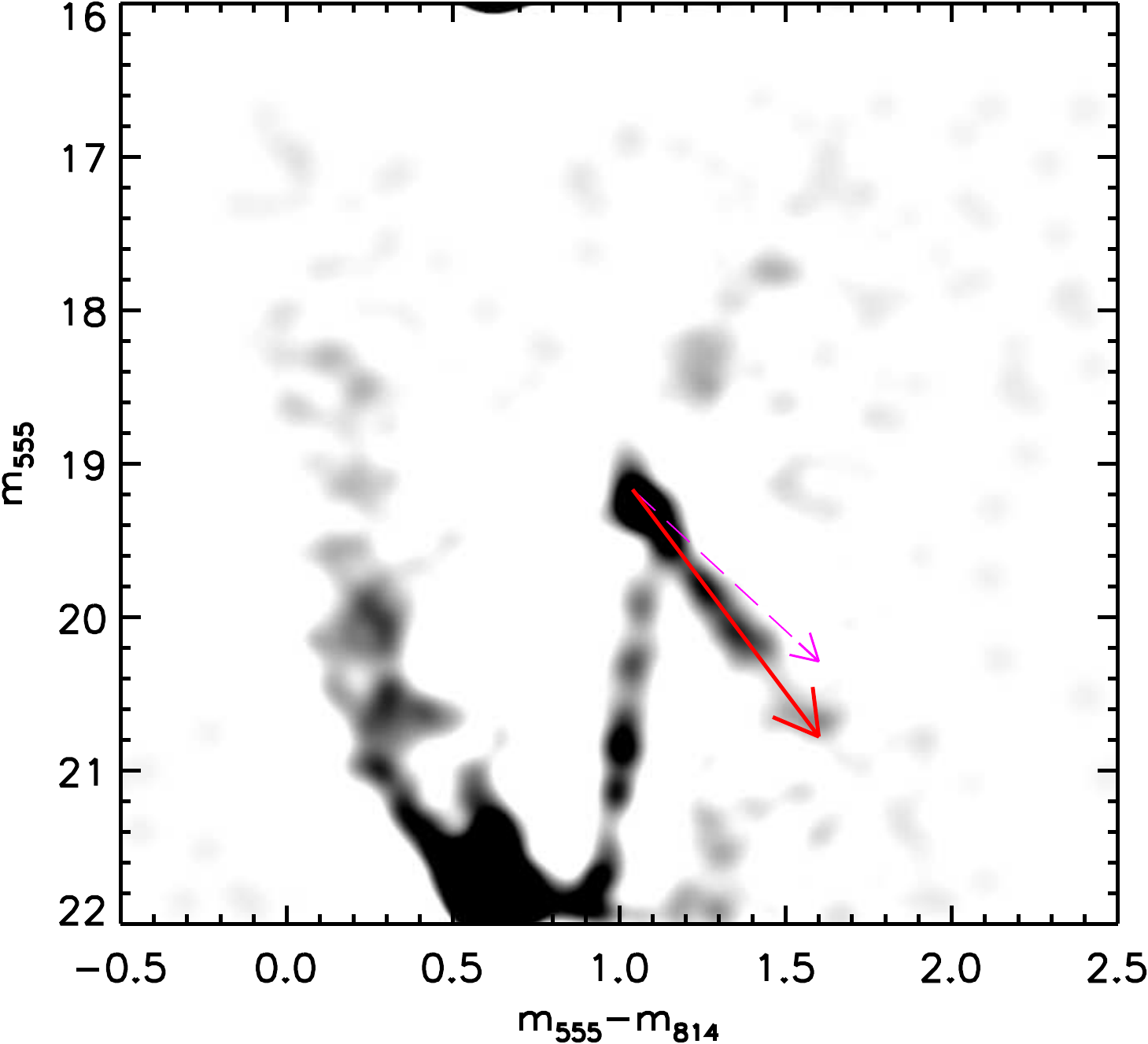}}
\caption{Same CMD as in Figure\,\ref{fig2}d after unsharp masking. The
thick arrow marks the ridge along the elongated RC, while the thin
dashed arrow shows the direction of the reddening vector in the diffuse
ISM, corresponding to $R_V=3.1$.} 
\label{fig3}
\end{figure}

The result of unsharp-masking the CMD of field stars is displayed in
Figure\,\ref{fig3}. The substructures and local density enhancements
appear better defined. Although all features were already present in
Figure\,\ref{fig2}d, the high density of more uniformly distributed
objects in their surroundings made them more difficult to distinguish
and characterise quantitatively. It is now possible to solidly determine
the ridge line of the extended RC and derive from it the direction of
the  reddening vector, without even having to know the location of the
nominal unextinguished RC, since the head of the distribution
empirically defines its position. The best linear fit along the extended
RC (thick arrow in Figure\,\ref{fig3}) provides the slope of the
reddening vector and is obtained by applying weights proportional to the
local object density. It is $\sim 1.5$ times steeper than the
reddening vector for the diffuse ISM in these bands (thin dashed arrow).
The slope along the extended RC corresponds to the ratio of absolute
($A$) and selective ($E$) extinction in the specific bands and its value
is

\begin{equation}
\frac{A_{555}}{E(m_{555}-m_{814})}=2.99\pm0.12.
\end{equation}

\noindent 
The corresponding slope in the $m_{814}, m_{555}-m_{814}$ plane is
$1.94\pm0.08$.

These slopes are in excellent agreement with those measured in the same
bands in the central regions of the 30\,Dor nebula (see De Marchi \&
Panagia 2014), respectively $2.97\pm0.08$ and $1.96 \pm 0.06$. The more
numerous set of filters available for those observations of 30\,Dor
allowed De Marchi \& Panagia (2014) to measure the extinction curve over
the wavelength range $0.3 - 1.6\,\mu$m and to determine the extinction
values at the wavelength corresponding also to the traditional Johnson
$B$ band. When the extinction properties are expressed in the form of
the ratio 

\begin{equation}
R_\lambda \equiv \frac{A_\lambda}{E(B-V)},
\end{equation}

\noindent where $A_\lambda$ is the extinction in the specific band and
$E(B-V)$ the colour excess in the Johnson $B$ and $V$ bands, the
resulting value for 30\,Dor is $R_V=4.5\pm0.2$. 

We note in passing that Merica--Jones et al. (2017) suggested that the
steeper slope of the extended RC in the CMD of 30\,Dor measured by  De
Marchi \& Panagia (2014), and the correspondingly large $R_V$ value,
might in fact result from the line-of-sight depth of the LMC, which
could in principle introduce an apparent grey contribution to the
extinction curve. Their models suggest that a depth of $5\pm1$\,kpc
(full width at half maximum) is required to account for the observed RC
shape in 30\,Dor, assuming an otherwise standard extinction law (i.e.
$R_V=3.1$). However, there are several problems with this
interpretation. {First, the measurements of the scale height of the
LMC along individual lines of sight do not support such a large depth 
(see above). Second, the morphology of the extended RC predicted by the
models of Merica--Jones et al. (2017; see their Figure\,7) bears no
resemblance to what is actually observed in the CMD of this field nor in
that of the 30\,Dor region (see also De Marchi et al. 2016). In
particular, our observations do not show the populous group of RC stars
predicted by those model to appear about $0.5$\,mag brighter than the
nominal location of the RC and caused by unextinguished RC stars closer
to the observer. In all our observations, the observed ''head'' of the
RC is exactly where theory of stellar evolution predicts it to be for
stars about $1-3$\,Gyr old at the nominal distance of the LMC (e.g.
Girardi \& Salari 2001; Salari \& Girardi 2002). Furthermore, if the
effects of differential reddening were compounded with those of
considerable line-of-sight depth, the RC would not develop along a 
straight line and would appear curved in the CMD. The fact that the
observations show a tight linear distribution instead indicates that the
effects of distance, if present, are marginal. }

Finally, and perhaps most importantly, the extinction law derived by De
Marchi \& Panagia (2014) from the morphology of the extended RC in
30\,Dor agrees very well with the one measured in the same field by
Ma{\'i}z--Apell\'aniz et al. (2014) with a completely independent
method. These authors obtained the extinction law from the observations
of 83 stars of spectral type O and B in the inner $2\arcmin$ radius of
30\,Dor, using spectroscopy and near infrared (NIR) photometry from the
VLT-FLAMES Tarantula Survey (Evans et al. 2011) as well as HST
photometry at optical wavelengths (De Marchi et al. 2011). Following a
Bayesian approach to derive the so-called extinction without standards
(Fitzpatrick \& Massa 2005), the { shape of the } extinction curve
that they obtain is in excellent agreement with that of De Marchi \&
Panagia { (2014; see their Figure\,4)} {at optical wavelengths
and shows a similar decay in the NIR, where De Marchi \& Panagia (2014)
measured the extinction curve directly from the HST observations while
Ma{\'i}z--Apell\'aniz et al. (2014) assumed it to be that of
Cardelli et al. (1989)}. The 50 stars with the smallest uncertainties on
extinction {in the sample of Ma{\'i}z--Apell\'aniz et al. (2014)
indicate $R_V=4.4\pm0.2$, fully matching the $R_V=4.5\pm0.2$ found by De
Marchi \& Panagia (2014). None of these curves is compatible with the
standard extinction law (i.e. $R_V=3.1$) suggested by Merica--Jones et
al. (2017).}

As discussed by De Marchi \& Panagia (2014; see also De Marchi et al.
2014, 2016), the extinction curve of 30\,Dor is fully compatible with
the contribution of two components: one being the extinction of the
standard diffuse ISM (e.g., Cardelli et al. 1989),
and the other an additional grey component accounting for about $1/3$ of
the total extinction. {The similarity between the slopes of the
reddening vectors measured in the F555W and F814W bands in the
NGC\,1938/1939 field and in 30\,Dor suggests that also in this region
the extinction is affected by an additional grey component at optical
wavelengths. Actually, with the extinction law of 30\,Dor (De Marchi \&
Panagia 2014), a value of $R_V=4.3\pm 0.3$ would result in a ratio
$A_{555}/E(m_{555}-m_{814})=2.99 \pm 0.12$, as measured in the 
NGC\,1938/1939 field. The grey component would account for about 30\,\%
of the total extinction and the diffuse ISM for the remaining $\sim
70$\,\% (since $R_V=3.1$ in the diffuse ISM). This } implies an
overabundance of big grains with respect to the Galactic and LMC ISM. In
the next section we will discuss the origin and implications of an
elevated value of $R_V$ in this region.

\section{Discussion}

A large value of $R_V$ has long been known to characterise the ISM in
star-forming regions in the Milky Way (MW; e.g., Baade \& Minkowski
1937; Watson \& Costero 2011 and references therein), so it does not
surprise that a large $R_V$ value is also found in 30\,Dor, where
massive star formation is currently ongoing (e.g. Walborn 1991). What
could appear unexpected, however, is to find an equally large value of
$R_V$ in the field containing NGC\,1938 and NGC\,1939, where the most
recent star-formation episode dates back to some 120\,Myr ago (see
Figure\,\ref{fig2}c). However, as we will show, this value of $R_V$
appears to be quantitatively consistent with the recent star formation
history of NGC\,1938.

Measuring $R_V=4.3$ in this field suggests that, like in the case of
30\,Dor, also here the extinction law is flatter than in the diffuse
Galactic ISM, where $R_V=3.1$ (e.g. Cardelli et al 1989). Such a
flatter, greyer extinction requires big grains to dominate the
distribution function of grain sizes (e.g. van de Hulst 1957). In the
case of 30\,Dor, from the study of the extinction curve at optical and
near infrared wavelength De Marchi \& Panagia (2014) concluded that the
relative abundance of big grains must be a factor of $2.2$ higher than
in the diffuse ISM. In principle, big grains could be produced through
mechanisms such as coalescence of small grains, small grain growth, and
selective destruction of small grains. However, from a study of the
ultraviolet extinction rproperties of three massive stars in 30\,Dor, De
Marchi \& Panagia (2019) showed that the excess of big grains does not
come at the expense of small grains, which are still at least as
abundant in 30\,Dor as in the diffuse ISM of the LMC. They concluded
instead that the dominant mechanism is fresh injection of big grains
with radius $a \ga 0.05\,\upmu$m, and that explosions of massive stars as
SNe II are a natural process to account for it, since massive star
formation is still ongoing in 30\,Dor (e.g. Walborn 1991; De Marchi et
al. 1993). 

Indeed, observations with ALMA and/or {\em Herschel} revealed up to
$\sim 0.5$\,\Msolar\ of dust and possibly more formed in situ in
core-collapse SNe ejecta, such as SN\,1987A (Matsuura et al. 2011; Gomez
et al. 2012; Indebetouw et al. 2014), the Crab Nebula, and Cassiopeia\,A
(Barlow et al. 2010; Gomez et al. 2012; De Looze et al. 2017). Although
the exact size distribution of dust formed in supernova remnants is
uncertain (e.g., Temim \& Dwek 2013; Owen \& Barlow 2015; Wesson et al.
2015; Bevan \& Barlow 2016), grains larger than $\sim 0.05\,\upmu$m
appear to be both produced during and preserved after the explosion of
SNe II (e.g. Gall et al. 2014; Silvia et al. 2010; Biscaro \& Cherchneff
2016). {Therefore, even though coalescence and growth of small
grains remain possible, SNe II explosions appear a likely and rather
natural source of big grains in regions that in the recent past
underwent massive star formation.}

For 30\,Dor, De Marchi \& Panagia (2014, 2019) discovered that at
wavelengths longer than the $V$ band the effects of big grains in the
extinction curve start to saturate and that beyond $\sim 1\,\upmu$m the
extinction curve tapers off as $\lambda^{-1.7}$, following almost
exactly the properties of the extinction law in the diffuse ISM (e.g.
Cardelli, Clayton \& Mathis 1989). These findings provide direct
information on the physical characteristics of the additional component
of big grains, including their size ($a \ga 0.05\,\upmu$m) and relative
abundance with respect to the diffuse ISM (a factor of about 2 larger).
The stark similarities between the reddening vectors measured in the
F555W and F814W bands in 30\,Dor and in NGC\,1938, as well as the
similar chemical composition, strongly suggest that the extinction law
in this cluster will remain comparable also at other wavelengths.
Therefore, as mentioned above, it is reasonable to assume that a
similar size and abundance of the big grains also apply to the field of
NGC\,1938.

In order to put this hypothesis on firmer observational grounds, we
explore whether the excess of big grains implied by $R_V=4.3$ as
measured in this field is consistent with that expected from SNe II
explosions in NGC\,1938. We start by calculating the mass of grains
along the line of sight.

In general, at short enough wavelengths the extinction cross section
$\sigma_{\rm ext}$ of a grain tends asymptotically to twice its
geometric cross section $\sigma_{\rm geom} = \pi a^2$, where $a$ is the
grain's radius (see, e.g., van de Hulst 1957; Greenberg 1968; Draine \&
Lee 1984). At longer wavelengths, where scattering becomes less
important and the extinction is dominated by pure absorption, the cross
section is smaller than $\sigma_{\rm geom}$ and approaches $\sigma_{\rm
geom} \times 2 \, \pi \, a/\lambda$. Conveniently enough, the transition
between the two regimes occurs approximately at $\lambda_0 \sim
2\,\pi\,a$. So in the ideal case of a fixed grain size, around
wavelength $\lambda_0$ one would expect a change of slope in the
extinction curve, with a steepening at longer wavelengths. Therefore,
adopting once again as representative of this field the extinction law
of 30\,Dor (De Marchi \& Panagia 2019), the saturation effect seen at
wavelengths longer than the $V$ band suggests that  $\lambda_0 \simeq
0.55\,\upmu$m and thus the presence of grains of radius $a\simeq
0.09\,\upmu$m or $9 \times 10^{-6}$\,cm. Their cross section
$\sigma_{\rm ext}$ is thus $2\,\pi\,a^2 = 5.1 \times 10^{-10}$\,cm$^{-2}$. 

\begin{figure}
\centering
\resizebox{\hsize}{!}{\includegraphics{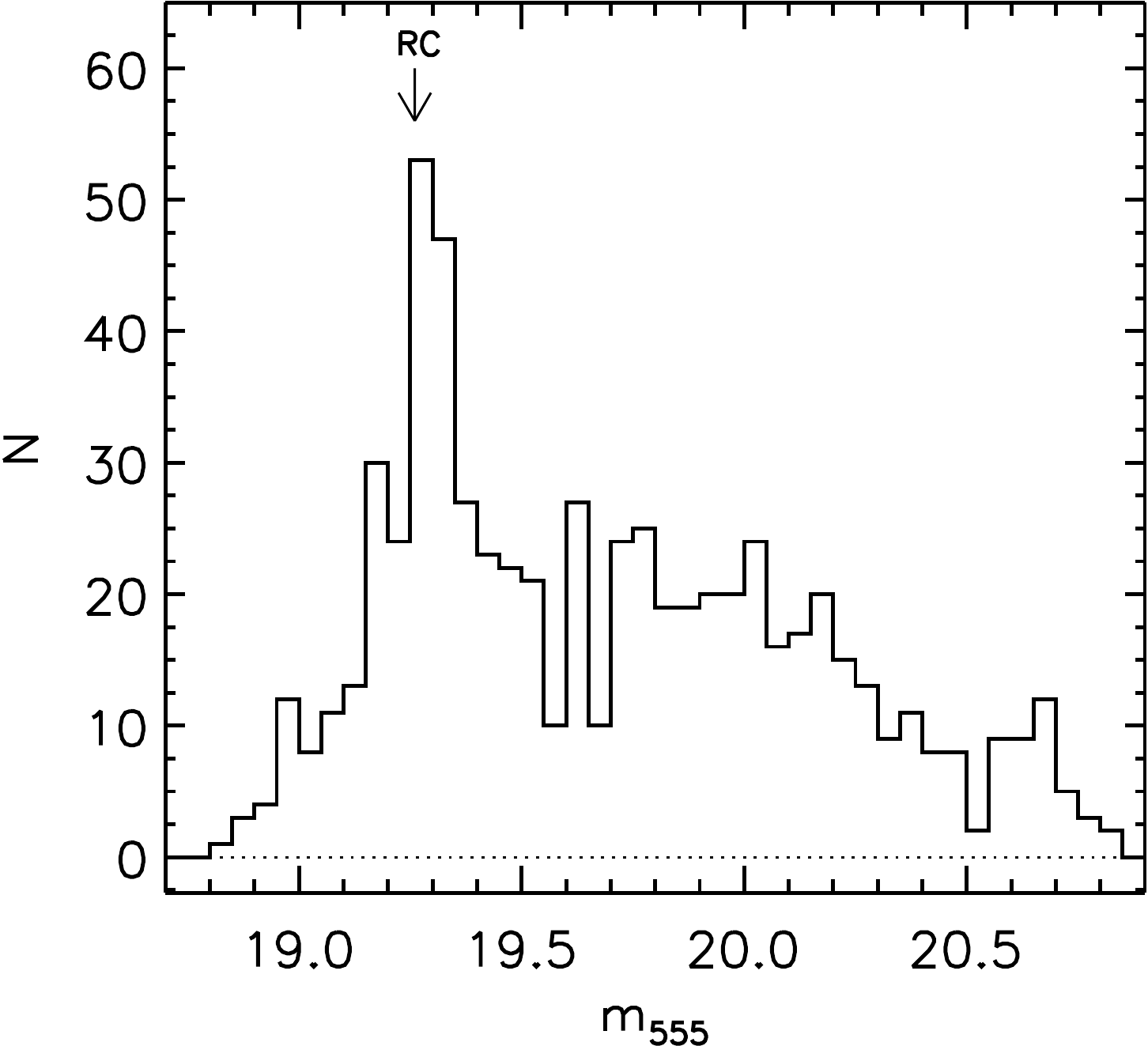}}
\caption{Histogram of the $m_{555}$ magnitudes of stars within a $\pm
0.4$\,mag  band around the reddening vector shown in Figure\,\ref{fig3}.
The peak at $m_{555}\simeq19.3$ agrees very well with the expected magnitude
of the `nominal' RC in this field (black arrow).}
\label{fig4}
\end{figure}

The number of such grains along the line of sight can be derived as the
ratio of the optical depth $\tau_V$ of the big grains component and
$\sigma_{\rm ext}$. As for $\tau_V$, it can be obtained directly from
the measured $A_V$ value pertaining just to the additional component of
big grains. The amount of $A_V$ in this field varies considerably, as
indicated by the extent of the RC. In Figure\,\ref{fig4} we display a
histogram of the $m_{555}$ magnitudes of stars within a $\pm 0.4$\,mag band
around the reddening vector shown in Figure\,\ref{fig3}. The peak in the
distribution  observed at $m_{555}\simeq19.3$ is in excellent agreement with
the magnitude of the `nominal' RC derived by De Marchi \& Panagia (2014)
for field stars in the 30\,Dor region, taking into account the somewhat
larger foreground extinction in this field ($A_V=0.34$; Mackey \&
Gilmore 2004) compared with the $A_V=0.22$ value applicable to the
30\,Dor area, as discussed in  Section\,3. Many RC stars are displaced
by reddening to fainter magnitudes by up to about $1.5$\,mag. The median
extinction value of RC stars is $A_V=0.36$ (see Figure\,\ref{fig4}),
which we will use as characteristic value for this field. We note that
the distribution of extinction values for RC stars around the $\sim
500$\,Myr old KMK88\,49 is in full agreement with that of the rest of
the field. But inside the NGC\,1938 cluster the median extinction value
is higher and $A_V=0.65$ is the amount indicated by the best fit to the
cluster MS (see Figure\,\ref{fig2}c), once the $A_V=0.34$ contribution
of the Galactic foreground extinction is removed. It is not surprising
that the extinction is higher in NGC\,1938 since, as we will show, it
appears that the cluster itself is the source of the elevated and
anomalous extinction in this field.

With a characteristic $A_V=0.36$ value in this field, the  fraction
contributed by big grains amounts to $A_V=0.10$, as we showed in
Section\,4 that the extra component of big grains account for about
30\,\% of the total extinction. In turn, the optical depth of big grains
is $\tau_V = 1.086 \times A_V = 0.11$ and the number of big grains along
the typical line of sight in this region is then $n=\tau_V/\sigma_{\rm
ext}=2.1 \times 10^8$. Considering that the area $A$ covered by the
observations is $50 \times 50$\,pc$^2$ or $2.4 \times 10^{40}$\,cm$^2$,
the total number $N$ of big grains in the region is $N=n\times
A=5.1\times 10^{48}$. With a typical density of $\sim 2.7$\,g\,cm$^{-3}$
(Panagia 1974; Draine \& Lee 1984), the mass of a typical grain is
$m=8.2 \times 10^{-15}$\,g. Hence the total mass of the extra component
of big  grains is $M=N \times m = 4.2 \times 10^{34}$\,g, or about
21\,\Msolar. 

If each SNe II explosion accounts for about $0.3$\,\Msolar\ of dust in
big grains (see above), about 70 such explosions would be needed over
the lifetime of NGC\,1938. Although the calculations are necessarily
approximate, this number is indeed fully compatible with the number of
SNe progenitors implied by the mass function of the cluster.

Using the 120\,Myr isochrone in Figure\,\ref{fig2}d as a reference,
within a radius of 22\arcsec or $5.5$\,pc of the centre of NGC\,1938
there are 86 stars with masses in the range $3 - 4$\,\Msolar. To remove
field-star contamination, we calculated the number of objects in the
same region of the CMD for an area of the same size located near the NE
edge of the image, finding on average 9 stars. This brings to 77 the
effective number of NGC\,1938 stars with masses in the range $3 -
4$\,\Msolar. These objects are all still on the MS (see
Figure\,\ref{fig2}c) and are bright enough to not suffer the effects of
photometric incompleteness. With 77 stars in this mass range, adopting a
standard IMF (Kroupa 2001) with power-law index $\gamma=-2.3$ for stars
above $0.5$\,\Msolar\ results in about $\sim 66$ SNe II progenitors, i.e.
stars initially more massive than $\sim 8$\,\Msolar. {Considering
the unavoidable approximations involved in this computation, the result
appears in surprisingly good agreement with the number of SNe II explosions
needed to account for the observed additional component of big grains in
the ISM of this region. Therefore, SNe II are a plausible source of the
additional big grains.}

The fact that the extinction is patchy means that the total amount of
dust along different lines of sight is not the same, but the tight
linear RC feature in the CMD suggests that the grains are well mixed,
with more or less similar distributions of grain sizes along
neighbouring lines of sight in this field. Hence the value of $R_V$
appears to be uniform across the area. This is probably not unexpected
if SNe II are at the source of the elevated value of $R_V$: at the age
of NGC\,1938 there should have been sufficient time for grains to mix
and be processed in the ISM since the first SNe II explosions some 100
Myr ago, at least over the field covered by our observations.

\section{Conclusions}

In summary, the flatter extinction curve implied by the $R_V=4.3$ value
measured in this field suggests that big grains are about twice as
abundant than in the diffuse ISM and this appears to be consistent with
the grain contributed by SNe II explosions throughout the life of
NGC\,1938. The initial total mass of the cluster, assuming a power-law
IMF index $\gamma=-1.3$ in the stellar mass range $0.08 - 0.5$\,\Msolar\
(Kroupa 2001), is about $4\,800$\,\Msolar. This places the cluster in
the intermediate mass range: about an order of magnitude smaller than
massive Galactic star-forming clusters such as Westerlund 1 ($\sim
50\,000$\,\Msolar; Andersen et al. 2017), Westerlund 2 ($\sim
36\,000$\,\Msolar; Zeidler et al. 2017), and NGC\,3603 ($\sim
15\,000$\,\Msolar; Harayama et al. 2008), NGC\,1938 is still a factor of
a few more massive than the Orion Nebula Cluster (ONC, $\sim
1\,000$\,\Msolar; Da Rio et al. 2012) 

An anomalous extinction curve (with $R_V \ga 5$) has long been known to
characterise the ONC (e.g. Baade \& Minkowski 1937; Sharpless 1952;
Johnson \& Mendoza 1964; Breger et al. 1981), and has been attributed to
grain growth in the dense prestellar molecular cloud through accretion 
from the gas phase or grain coagulation (e.g. Cardelli \& Clayton 1988).
Other examples of star-forming regions in dense molecular clouds with 
anomalous extinction are Taurus (Whittet et al. 2000) and Cep\,OB3b
(Allen et al. 2014). While NGC\,1938 is about 5 times more massive than
the ONC, it has likely formed in an equally dense molecular cloud. What
is surprising, however, is that the excess of  big pre-stellar grains
persists some $\sim 120$\,Myr after the cluster's formation: by now the
molecular cloud has long completely dissipated, thereby removing the
conditions for grains to grow. If instead the origin of the
overabundance of big grains in NGC\,1938 is the fresh injection of big
grains by SNe II explosions, as we suggest, it appears easier to
understand an anomalous extinction curve well past the end of the star
formation process: SNe II explosions happen after the formation of the
cluster and continue to replenish the ISM with fresh new grains for up
to $\sim 40$\,Myr, when all SNe II progenitors have exploded (e.g.
Marigo et al. 2017). Once injected, the additional big grains will
continue to pollute the ISM for a non-negligible amount of time. A study
of a complete sample of SN remnants in the Magellanic clouds by Temim et
al. (2015) estimates grain lifetimes in the range $\sim 20 - 70$\,Myr,
with $\sim 50$\,\% uncertainties. Big grains from SN II explosions are 
thus not unexpected even in the $\sim 120$\,Myr old NGC\,1938.

Therefore, an anomalous extinction could be a common and long-lasting
characteristic of star forming regions that are sufficiently massive to
host SNe II. An intermediate-mass cluster such as NGC\,1938 appears to
be able to considerably alter the conditions of the surrounding ISM. 
{Indeed, in a forthcoming paper (De Marchi et al., in prep.) we will
show that the LMC regions with a prominent extended RC and apparently
anomalous extinction (including also relatively young clusters such as
NGC\,1756, NGC\,1858, NGC\,1872, and NGC\,1903) are all characterised by
extended dust emission detected in the far infrared, which is not
present in regions of low total extinction. }

{Of course, the anomalous extinction caused by the pollution by
large grains will affect the observations of all regions in the
background. This has important consequences for our ability to derive
meaningful values of the mass or the star formation rates from the study
of extragalactic stellar populations, regardless of their age.} Normally
extinction correction is needed in order to derive the intrinsic
luminosity of the stars. To this aim, the extinction properties in the
region must be known, including the value of $R_V$. This requires a
careful analysis of the stellar populations, including those in
surrounding regions, to identify signs of intermediate-mass clusters
formed in the past few 100\,Myr that could act as potential ISM
polluters. This is not always possible. For instance, with a current
total mass of about $3\,300$\,\Msolar, a cluster like NGC\,1938 is easy
to identify in the Magellanic clouds, but already at the distance of
M\,31 it would be problematic to recognise clusters with the mass of the
ONC at ages $\la 200$\,Myr (Johnson et al. 2015), due to detection and
resolution limits. At larger distances it becomes increasingly difficult
to identify even more massive or younger clusters, and a larger
uncertainty on the population's physical parameters as derived from
photometry and line luminosities cannot be avoided. Paradoxically, the
situation might become less uncertain in the early universe if the
Magellanic Clouds are indeed representative of these low-metallicity
primeval environments (e.g. Simon et al. 2007). In that case, since
young unresolved massive star-forming regions are the only objects
detectable at high redshift, our work suggests that a value of $R_V$ not
smaller than $\simeq 4.5$ appears most likely. 

\vspace*{0.5cm}

We are grateful to an anonymous referee whose comments have helped us
improve the presentation of this work. APM acknowledges support from the
European Research Council (ERC) under the European Union's Horizon 2020
research innovation programme (Grant Agreement ERC-StG 2016, No 716082
'GALFOR', http://progetti.dfa.unipd.it/GALFOR), by the MIUR through the
FARE project R164RM93XW 'SEMPLICE' and the PRIN programme 2017Z2HSMF.

\end{document}